\documentclass[amsmath,amssymb,iopams]{iopart}

\usepackage{graphicx}

\newcommand{\Na}[1]{{\hat{#1}}}
\newcommand{\Fe}[1]{{\bi #1}_{\rm F}}
\newcommand{\Fep}[1]{{\bi #1}'_{\rm F}}
\newcommand{\Avp}[1]{\left\langle{#1}\right\rangle_{\Fe p}}
\newcommand{\Avpp}[1]{\left\langle{#1}\right\rangle_{\Fep p}}

\begin{document}

\title[Superconductor-Ferromagnet Bi-Layers]{Superconductor-Ferromagnet
Bi-Layers: a Comparison of $s$-Wave and $d$-Wave Order Parameters}

\author{T L\"uck and U Eckern}

\address{Institut f\"ur Physik, Universit\"at Augsburg, 86135 Augsburg, Germany}
\ead{lueckt@physik.uni-augsburg.de}

\begin{abstract}
  We study superconductor-ferromagnet bi-layers, not only for $s$-wave
  but also for $d$-wave superconductors. We observe oscillations of
  the critical temperature when varying the thickness of the
  ferromagnetic layer for both $s$-wave and $d$-wave superconductors.
  However, for a rotated $d$-wave order parameter the critical
  temperature differs considerably from that for the unrotated case.
  In addition we calculate the density of states for different
  thicknesses of the ferromagnetic layer; the results reflect the
  oscillatory behaviour of the superconducting correlations.  
\end{abstract}

\section{Introduction}

The interplay of superconductivity and ferromagnetism has been studied
for many years~\cite{FF64,LO64}. Nowadays the focus is on
hybrid-structures of superconductors and various ferromagnetic
materials.  For bi-layers consisting of a superconductor and a
ferromagnetic metal an oscillation of the critical temperature has
been found -- experimentally~\cite{Mu96,Ru02,Si03} as well as
theoretically~\cite{Kh97,Ta98} -- when increasing the thickness of the
ferromagnetic layer. Similar observations have been made for
superconductor-ferromagnet multi-layers~\cite{Rad91,Kh97}. For two
superconductors which are coupled via a thin layer of a ferromagnetic
metal an oscillation of the critical current when varying the layer
thickness has also been reported~\cite{Ry01,Ko02}; depending on the
layer thickness such systems can be $\pi$-junctions.  Recently a layer
of a ferromagnetic metal attached to a bulk superconductor has been
considered, resulting in the density of states showing
oscillations~\cite{Bu00,Za01,Za02,Ko01}, too.  The origin of these
effects is a state in the ferromagnetic metal which is similar to that
proposed for a ferromagnetic superconductor by Larkin and
Ovchinnikov~\cite{LO64} and Fulde and Ferrell~\cite{FF64}: in the
presence of a spin exchange field the superconducting order parameter
is spatially oscillating. A LOFF-like state can be present in a
ferromagnet where superconducting correlations enter via the proximity
effect~\cite{Ta98}.

\begin{figure}[b]\begin{center}
\includegraphics[scale=0.5]{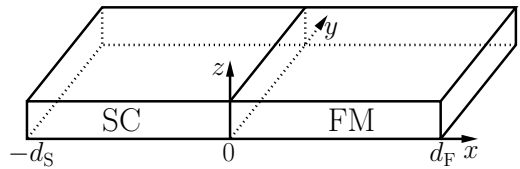}\end{center}
\caption{\label{fig1}A bi-layer consisting of a superconductor
  of thickness $d_{\rm S}$ and a ferromagnetic metal of thickness
  $d_{\rm F}$. The interface between both materials (at $x=0$) is
  assumed to be completely transparent, and the sample is
  translationally invariant in $y$- and $z$-direction.}
\end{figure}

Up to now only $s$-wave superconductors have been studied.  In view of
possible applications, it is also important to consider an order
parameter with $d$-wave symmetry, which presumably is realized in the
cuprates; there the crystallographic orientation fixes the direction
of the $d$-wave lobes. For a related experimental investigation
see~\cite{Fr03}.

In this work we extend the theory to $d$-wave superconductors, where
we have to account for the anisotropy of the order parameter.  This
means that in the quasi-classical framework, which we use, the
Usadel equation is no longer applicable, and we have to use the
Eilenberger equation instead.  It should be mentioned that the
Eilenberger equation is also needed for the description of $s$-wave
superconductors in the clean limit, which was pointed out for the
current problem in~\cite{Be02}.

In the following we study a bi-layer of a superconductor and a
ferromagnetic metal as presented in \fref{fig1}. As the structures we
are interested in are three-dimensional our mean field approach is of
sufficient accuracy. We consider the behaviour of the critical
temperature and of the density of states.  In particular we compare
the results for different order parameter symmetries, namely of the
$s$-wave and $d$-wave type. First we briefly introduce the fundamental
quasi-classical equations.  Afterwards we present the results, and we
finish with a short conclusion.

\section{Method}
To study superconductors in the vicinity of boundaries we apply the
theory of quasi-classical Green's functions in thermal
equilibrium~\cite{Ei68,RaSm86}. The Green's functions are determined
from the Eilenberger equation
\begin{eqnarray}\label{Eilen_eq}
\big [\Na\tau_3E+
I({\bi r})\Na\sigma_3+i\Na\Delta(\Fe p,{\bi r})-\Na\Sigma({\bi r}),
\Na g(E,\Fe p;{\bi r})\big ]+\nonumber\\
+i\Fe v\cdot\partial_{\bi r}\Na g(E,\Fe p;{\bi r})=0\;.
\end{eqnarray}
and must fulfil the normalisation condition
\begin{equation}\label{norm}
\left[\Na g(E,\Fe p;{\bi r})\right]^2=\Na 1\;.
\end{equation}
Here $\Na\tau_i$ is the direct product of the $i^{\rm th}$
Pauli-matrix in Nambu space and the identity in spin space; vice versa
$\Na\sigma_i$ is the direct product of the $i^{\rm th}$ Pauli-matrix
in spin space and the identity in Nambu space.  Consequently the
Green's function, $\Na g$, has a $4\times 4$ matrix structure. For our
purpose it is sufficient to choose the orientation of the internal
spin-exchange field of the ferromagnetic metal in $z$-direction, ${\bi
  I}({\bi r})=I({\bi r}){\bi e}_z$, which leads to the term $I({\bi
  r})\Na\sigma_3$ in the Eilenberger equation.

The superconducting order parameter, which we assume to be
spin-singlet, reads
\begin{equation}
\hat \Delta(\Fe p,{\bi r})=
\pmatrix{0 & i\Na\sigma_2\Delta(\Fe p,{\bi r})\cr -i\Na\sigma_2\Delta^*(\Fe p,{\bi r})& 0}
\end{equation}
with
\begin{equation}
\Delta(\Fe p,{\bi r})=\eta(\Fe p)\Delta({\bi r}).
\end{equation}
The symmetry of the order parameter is determined by the basis
function, $\eta(\Fe p)$:
\begin{equation}
\eta(\Fe p)=
\cases
{1 &  $s$-wave($\eta_s$)\cr
(p_{{\rm F}x}^2-p_{{\rm F}y}^2)/p_{\rm F}^2 & unrotated $d$-wave ($\eta_d$)\cr
p_{{\rm F}x}p_{{\rm F}y}/p_{\rm F}^2 & $45^\circ${-rotated} $d${-wave} ($\eta'_d$)}.
\end{equation}
The order parameter must be determined self-consistently via
\begin{equation}\label{OP}
\Na\Delta({\bi r})=-\pi V{\cal N}_0 T\sum_{|E_n|<E_c}
\frac{\Na\sigma_2}{2}{\rm Tr}_\sigma \left[\Na\sigma_2
\Avp{\eta(\Fep p)\Na g(E,\Fe p;{\bi r})}\right]\;,
\end{equation}
where $\Avpp{\dots}$ denotes an average over the Fermi surface, which
is assumed to be spherical in the $s$-wave case and cylindrical in the
$d$-wave case (this is justified for the layered cuprate
superconductors).  The cut-off energy is $E_c$, the attractive
interaction is $V<0$, and the normal density of states per spin at the
Fermi energy is denoted by ${\cal N}_0$.  Impurity scattering is
treated in Born approximation which leads to the following
self-energy:
\begin{equation}
\Na\Sigma(E,{\bi r})=\frac{-i}{2\tau}
\Avp{\Na g(E,\Fe p;{\bi r})}
\end{equation}
with the scattering time $\tau$.

It is important to note that in principle the quasi-classical theory is
only valid if all energy scales are small compared to the Fermi
energy, $E_{\rm F}$; for most superconductors this is the case as $T_c\ll
E_{\rm F}$.  However, for many ferromagnetic materials the exchange energy
is of the same order of magnitude as $E_{\rm F}$. Strictly speaking this
theory can therefore only be applied for rather weak ferromagnets.
Strong ferromagnets have been treated in some special cases:
superconductors in proximity to half-metals (where only one
spin-channel is metallic)~\cite{Es03} or in contact to strong
ferromagnetic insulators~\cite{To88,Fo00} have been examined by
extensions of the quasi-classical theory.

\section{Results}

First we investigate the oscillations of the pairing function in a
quite simple system which we can treat analytically: we consider a
ferromagnetic layer of thickness $d_{\rm F}$ attached to a bulk
superconductor ($d_{\rm S}\to\infty$, see \fref{fig1}) without disorder;
for simplicity we consider the case of a spherical Fermi surface in
both materials, and an identical Fermi velocity, $v_{\rm F}$.  Furthermore
we assume a completely transparent interface at $x=0$, a specular
surface at $x=d_{\rm F}$, and a spatially constant order parameter in the
superconductor, $\Delta(\Fe p,{\bi r})=\Delta(\Fe p)$.  Then the
normal part of the Green's function in the ferromagnetic metal reads
\begin{equation}\label{g_diag}
g_{\uparrow\uparrow/\downarrow\downarrow}(E,\Fe p)=
\frac{1-{\rm e}^{i\vartheta_{\uparrow/\downarrow}(E,\Fe p)}\alpha(E,\Fe p)\beta(E,\Fep p)}
{1+{\rm e}^{i\vartheta_{\uparrow/\downarrow}(E,\Fe p)}\alpha(E,\Fe p)\beta(E,\Fep p)}
\end{equation}
with
\begin{equation}
\alpha(E,\Fe p)=\frac{E-\sqrt{E^2-|\Delta(\Fe p)|^2}}{\Delta^*(\Fe p)},
\end{equation}
\begin{equation}
\beta(E,\Fe p)=
-\frac{E-\sqrt{E^2-|\Delta(\Fe p)|^2}}{\Delta(\Fe p)},
\end{equation}
and
\begin{equation}\label{spin_rot}
\vartheta_{\uparrow/\downarrow}(E,\Fe p)=\frac{2(E\mp I)\lambda(\Fe p)}{v_{\rm F}};
\end{equation}
$\lambda(\Fe p)=2d_{\rm F}p_{\rm F}/p_{F}^{||}$ ($p_{F}^{||}$: Fermi momentum
parallel to the $y$-$z$-plane) is the length of the classical
trajectory in the ferromagnetic layer. Note that $\Fep p$ is uniquely
determined by $\Fe p$ for a specular surface since the parallel
momentum is conserved, $p_{F}^{||}={p'_{F}}^{||}$.

The angle-resolved density of states in the ferromagnetic layer can be
expressed in terms of the normal part of the Green's function,
\begin{equation}
{\cal N}(E,\Fe p)=\frac{1}{2}{\cal N}_0{\rm Re}
\left[g_{\uparrow\uparrow}(E,\Fe p)+g_{\downarrow\downarrow}(E,\Fe p)\right].
\end{equation}

\begin{figure}[t]
\begin{center}\includegraphics{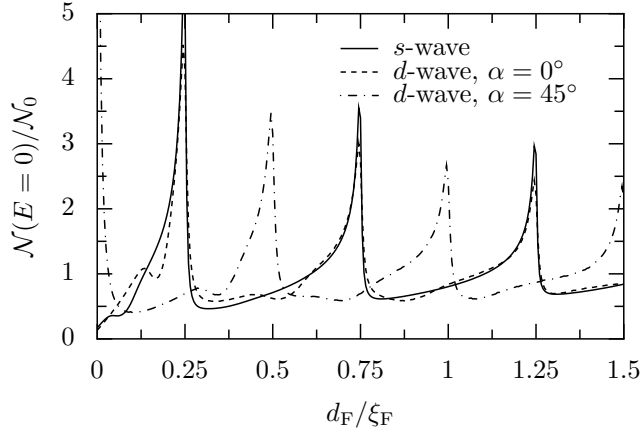}\end{center}
\caption{\label{fig2}The density of states at $E=0$ is shown as a function
  of the ferromagnetic layer thickness with $\xi_{\rm F}=\pi v_{\rm F}/I$. The
  $s$-wave and the unrotated $d$-wave case are almost identical with
  maxima at $d_{\rm F}=(1/2+k)\xi_{\rm F}/2$, where $k$ is an integer; for a
  rotated $d$-wave order parameter the maxima are shifted to
  $d_{\rm F}=k\xi_{\rm F}/2$. For the plot we added a finite imaginary
  part to the energy ($E\to E+i0.01T_c$).}
\end{figure}

We now consider the angle-averaged density of states at $E=0$ as a
function of $d_{\rm F}$, which is normalised by the ferromagnetic length,
$\xi_{\rm F}=\pi v_{\rm F}/I$.  In \fref{fig2} we present the results for the
$s$-wave case ($\Delta(\Fe p)=1.76T_{c0}$) as well as for the
unrotated ($\Delta(\Fe p)=2.14T_{c0}\eta_d(\Fe p)$) and for the
$45^\circ$-rotated $d$-wave cases ($\Delta(\Fe
p)=2.14T_{c0}\eta'_d(\Fe p)$); we choose an exchange field of
$I=10T_{c,0}$.

The zero-energy density of states for an $s$-wave and an unrotated
$d$-wave order parameter behaves quite similar up to minor deviations:
we find maxima at $d_{\rm F}=(1/2+k)\xi_{\rm F}/2$, $k\in {\mathbf
  N}_0$.  At these values of $d_{\rm F}$ Andreev bound states with a
large spectral weight exist in the gap region; i.e. in the gap region
the density of states is enhanced compared to the normal state value.
For a $45^\circ$-rotated $d$-wave order parameter we find maxima at
$d_{\rm F}=n\xi_{\rm F}/2$. The reason for this shift is that the
quasi-particles acquire an additional phase due to the scattering at
the surface which changes the sign of the order parameter. In
particular for $d_{\rm F}=0$ the commonly known zero-energy bound
states occur at surfaces~\cite{TL01}.

The density of states has been studied before~\cite{Za01,Za02,Za03}
including surface roughness and a finite transparency of the
superconductor-ferromagnet interface. The oscillatory behaviour has
also been observed experimentally~\cite{Ko01,Fr03}. A discussion of the
non-magnetic case can be found in~\cite{TL03}.

\begin{figure}[t]
\begin{center}\includegraphics{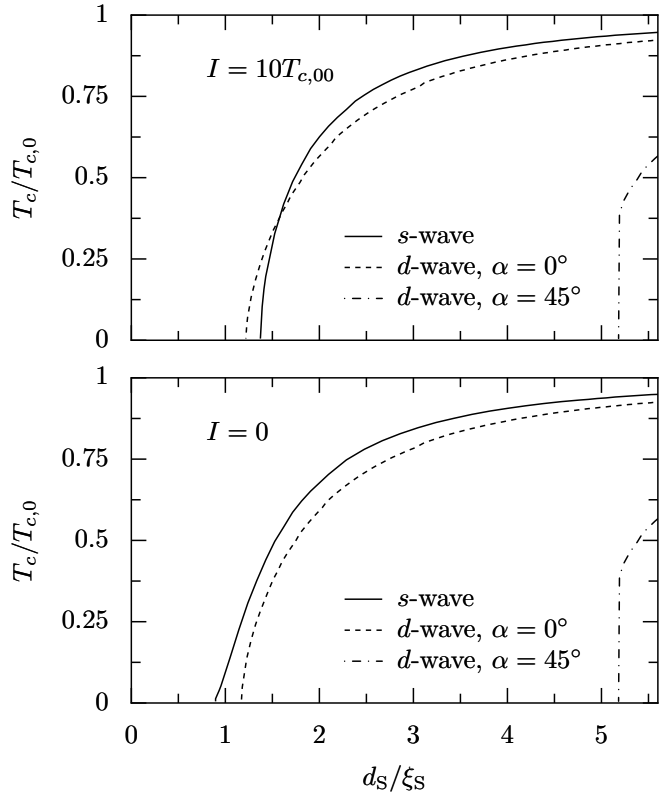}\end{center}
\caption{\label{Tc-dS}The critical temperature of a thin superconducting
  layer on top of a bulk ferromagnet ($I=10T_{c,00}$, $d_{\rm
    F}\to\infty$) as a function of the layer thickness, $d_{\rm S}$,
  which is normalised by the superconducting coherence length
  $\xi_{\rm S}$ (upper panel). For comparison the critical temperature
  is also shown for $I=0$ (lower panel).}
\end{figure}
\begin{figure}[t]
\begin{center}\includegraphics{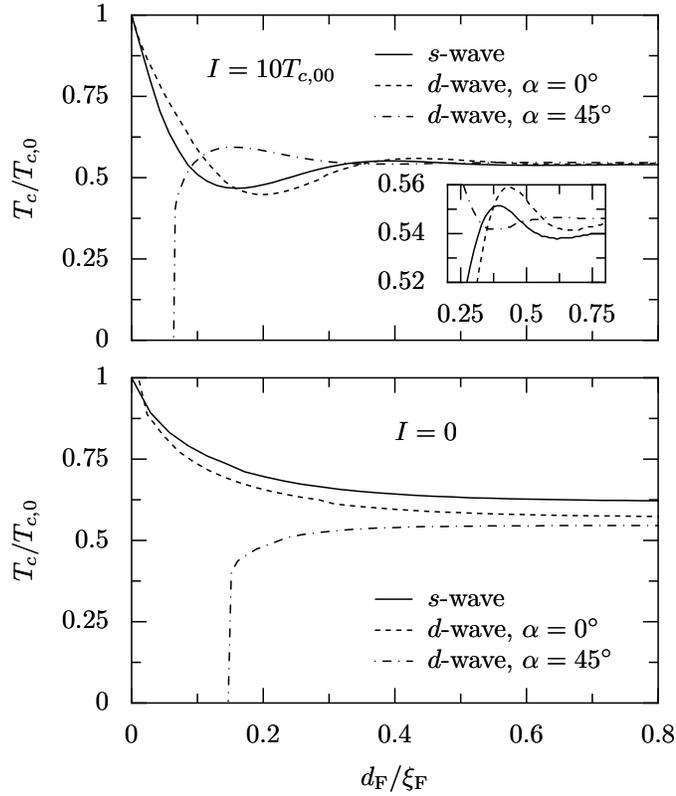}\end{center}
\caption{\label{Tc-dF}The critical temperature of a bi-layer with fixed
  $d_{\rm S}$ as a function of the ferromagnetic layer thickness
  $d_{\rm F}$ (upper panel). For all order parameter types $T_c$ is
  oscillating. The parameter $d_{\rm S}$ is chosen such that
  $T_c=0.54T_{c,0}$ for $d_{\rm F}\to\infty$; i.e. $d_{\rm S}$ equals
  $1.72\xi_{\rm S}$,$1.22\xi_{\rm S}$, and $5.52\xi_{\rm S}$ for the
  $s$-wave, unrotated and rotated $d$-wave situations, respectively.
  For comparison the critical temperature is also shown for the
  unrotated case, $I=0$, where no oscillations occur (lower panel).}
\end{figure}

\begin{figure}[t]
\begin{center}\includegraphics{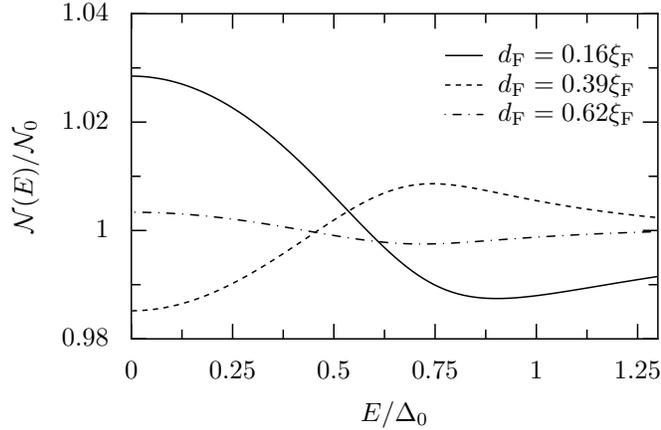}\end{center}
\caption{\label{doss} The density of states at $x=d_{\rm F}$ for an $s$-wave
  symmetry of the order parameter. These values of $d_{\rm F}$ correspond to
  the first two minima ($d_{\rm F}=0.16\xi_{\rm F},0.62\xi_{\rm F}$) and the
  first maximum ($d_{\rm F}=0.39\xi_{\rm F}$) of the critical temperature.}
\end{figure}
\begin{figure}[t]
\begin{center}\includegraphics{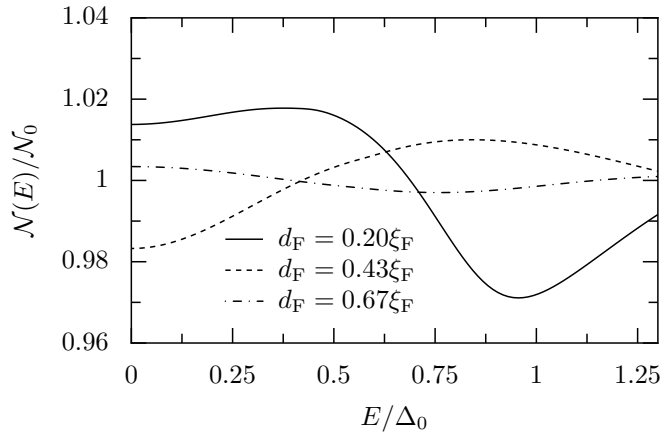}\end{center}
\caption{\label{dosd0} The density of states at $x=d_{\rm F}$ for an unrotated
  $d$-wave symmetry of the order parameter. These values of $d_{\rm F}$
  correspond to the first two minima ($d_{\rm F}=0.20,0.67\xi_{\rm F}$) and the
  first maximum ($d_{\rm F}=0.43\xi_{\rm F}$) of the critical temperature.}
\end{figure}

\begin{figure}[t]
\begin{center}\includegraphics{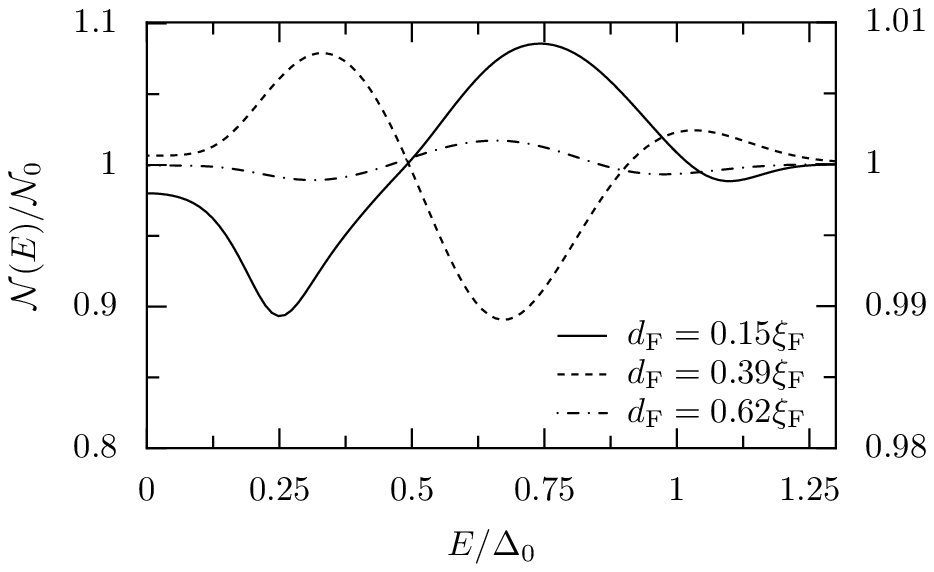}\end{center}
\caption{\label{dosd45} The density of states at $x=d_{\rm F}$ for a
  $45^\circ$-rotated $d$-wave symmetry of the order parameter.  The
  left axis applies to the density of states at the first maximum of
  $T_c$ ($d_{\rm F}=0.15\xi_{\rm F}$), whereas the right axis applies
  for the first minimum ($d_{\rm F}=0.39\xi_{\rm F}$) and the second
  maximum ($d_{\rm F}=0.62\xi_{\rm F}$).}
\end{figure}

In the following we focus on the critical temperature of bi-layer
systems.  As before we assume a completely transparent interface
between the ferromagnet and the superconductor; furthermore the Fermi
surfaces are supposed to be identical in both materials. The ferromagnet
is described by the exchange energy, $I$, and the impurity scattering
strength, $1/2\tau_{\rm F}$; we choose these parameters to be $I=10T_{c,0}$
and $1/2\tau_{\rm F}=5T_{c,0}$. This is a reasonable choice having in mind
ferromagnetic metals like Fe or Ni which are well-described by the
relations $T_c\ll I$, and $1/2\tau_{\rm F}\ll I$.

First we study the critical temperature of the system with an $s$-wave
superconductor, which we assume to be dirty, $1/2\tau_{\rm
  S}=10T_{c,0}$.  The coherence length at zero temperature, $\xi_{\rm
  S}$, is given by $\xi_{\rm S}=\sqrt{\xi_0 l_{\rm S}}\approx 0.53\xi_0$
where $\xi_0=v_{\rm F}/\Delta_0\pi$ ($\Delta_0=1.768T_{c,0}$) is the
BCS-value of the coherence length for a clean superconductor, and
$l_{\rm S}=\tau_{\rm S} v_{\rm F}$ is the mean free path.  In \fref{Tc-dS} we
present the critical temperature, $T_c$, for a superconducting layer
on a bulk ferromagnet ($d_{\rm F}\to\infty$) as a function of its
thickness, $d_{\rm S}$.  The critical temperature decreases with the
thickness of the superconductor, and below a critical layer thickness,
$d_{\rm S}=1.37\xi_{\rm S}$, the superconductivity vanishes. Now we fix
$d_{\rm S}=1.72\xi_{\rm S}$, for which $T_c=0.54T_{c,0}$ when $d_{\rm
  F}\to\infty$; for this thickness of the superconducting layer we
examine the critical temperature as a function of the ferromagnetic
layer thickness, $d_{\rm F}$.  We find an oscillation of the critical
temperature when varying the thickness of the ferromagnet (see
\fref{Tc-dF}).

The oscillations can be explained as follows: since no current can
flow across the surface at $x=d_{\rm F}$, the pairing function has to
obey particular boundary conditions, i.e. its derivative normal to the
surface has to vanish. In the presence of a spin exchange field the
pairing is spatially oscillating with wave length $\xi_{\rm F}=\pi
v_{\rm F}/I$~\cite{Ta98}. At the maxima of $T_c$ the thickness of the
ferromagnetic layer, $d_{\rm F}$, is such that the boundary conditions
are fulfilled quite naturally, whereas minima occur if the pairing
function has to be suppressed considerably to fulfil the boundary
conditions.  Therefore the distance of two neighbouring minima is
expected to be of the order of $0.5\xi_{\rm F}$, which is the same
periodicity as observed before for the density of states (see
\fref{fig2}). In our numerical calculation the first two minima of
$T_c$ can be found at $d_{\rm F}=0.16\xi_{\rm F}$ and $d_{\rm
  F}=0.62\xi_{\rm F}$ ($s$-wave case, see inset in \fref{Tc-dF}).
Their distance is $0.46\xi_{\rm F}$ which is close to the expected
value.  These results are consistent with other theoretical
findings~\cite{Kh97,Ta98} which could also be fitted to experimental
observations~\cite{Mu96,Si03}. For comparison we also present the
critical temperature for a non-magnetic metal layer ($I=0$); as
expected superconductivity is suppressed less effectively without
magnetism (see \fref{Tc-dS}). For this case, no oscillations with the
metal layer thickness are observed (see \fref{Tc-dF}), which is an
obvious result considering the discussion in relation
with~\eref{g_diag}-\eref{spin_rot}.

As previously discussed, the oscillating behaviour of the pairing function
can also be observed in the local density of states at $x=d_{\rm F}$. We
calculate the density of states for those values of $d_{\rm F}$ where the
critical temperature has a maximum or a minimum.  The results are
presented in \fref{doss}.  For the minima of $T_c$ (at
$d_{\rm F}=0.16,0.62\xi_{\rm F}$) the density of states in the gap region is
enhanced compared to the normal state; this is related to sub-gap
Andreev bound states which exist inside the ferromagnet. At the
maximum of $T_c$ ($d_{\rm F}=0.39\xi_{\rm F}$) we find a gap-like structure of the
density of states; i.e. in the gap region the density of states is
smaller than the normal state value. Altogether we find that bound
states in the gap region tend to suppress superconductivity, and can
be related to minima of $T_c$.

Now we will turn to the case of $d$-wave superconductors.  For a
$d$-wave symmetry of the order parameter the bulk value of the
critical temperature, $T_{c,0}$, is suppressed by non-magnetic
impurities, and $T_c$ is given by~\cite{HiGo93}
\begin{equation}
\ln\left(\frac{T_{c,00}}{T_{c,0}}\right)=
\Psi\left(\frac{1}{2}+\frac{1}{4\pi T_{c,0}\tau_{\rm S}}\right)-
\Psi\left(\frac{1}{2}\right)
\end{equation}
where $T_{c,00}$ is the critical temperature for a clean sample and
$\Psi(x)$ is the digamma function.  In the following we choose the
impurity scattering inside the superconductor to be
$1/2\tau_{\rm S}=0.1T_{c,00}$ which leads to $T_{c,0}=0.92T_{c,00}$; the
bulk order parameter at zero temperature is $\Delta_0=2.02T_{c,00}$
which is smaller than its value in the clean case
($\Delta_{00}=2.14T_{c,00}$). As the disorder is small the
superconducting coherence length is given by
$\xi_{\rm S}\approx\xi_0=v_{\rm F}/\Delta_0\pi$.

It is well-known that the behaviour of $d$-wave superconductors at
boundaries depends crucially on the orientation of the order parameter
with respect to the boundary. Therefore we compare the case where the
order parameter is rotated by $45^\circ$ with the unrotated case. 

An unrotated $d$-wave superconductor is expected to behave similar to
an $s$-wave superconductor. The reason is that along the classical
trajectories the order parameter does not change its phase. And indeed
the behaviour of an unrotated $d$-wave superconductor and an $s$-wave
superconductor is quite similar: the superconductivity of a thin layer
on a bulk ferromagnet is suppressed completely when its thickness is
below $d_{\rm S}=1.22\xi_{\rm S}$ (see \fref{Tc-dS}). If we fix the
superconducting layer thickness to $d_{\rm S}=1.94\xi_{\rm S}$ (which
leads to $T_c=0.54T_{c,0}$ for $d_{\rm F}\to\infty$) and vary $d_{\rm
  F}$ we find an oscillating behaviour of $T_c$ as before. The first
two minima are at $d_{\rm F}=0.20\xi_{\rm F}$ and $d_{\rm
  F}=0.67\xi_{\rm F}$, and the first maximum can be found at $d_{\rm
  F}=0.43\xi_{\rm F}$ (see \fref{Tc-dF}); the distance between the
first two minima is $0.47\xi_{\rm F}$.  Of course some quantitative
differences exist which are mainly due to the nodes of the $d$-wave
order parameter. The critical temperature for a non-magnetic metal
layer ($I=0$) behaves similar to the $s$-wave case: for an infinite
metal layer $T_c$ lies above the value for the magnetic case (see
\fref{Tc-dS}); the oscillations with a varying thickness of the
non-magnetic metal vanish (see \fref{Tc-dF}).

For those values of $d_{\rm F}$ which are related to a maximum or a minimum
of $T_c$ the density of states shows qualitatively the same behaviour
as for the $s$-wave case; i.e. for small energies the density of
states at minima of $T_c$ is enhanced compared to the normal state
value whereas a gap-like structure can be observed at the maxima of
$T_c$ (see \fref{dosd0}).

The situation changes drastically if the $d$-wave order parameter is
rotated by $45^\circ$: Surfaces are now pair-breaking as the
quasi-particles are scattered so that a sign change of the order
parameter occurs along their trajectories~\cite{TL01}; and it is
well-known that at specular surfaces the order parameter vanishes. As
a consequence the suppression of superconductivity is much stronger
than for the cases discussed before.  For a superconducting layer
which is on top of a bulk ferromagnet this can be seen in
\fref{Tc-dS}; in particular the critical thickness of the
superconductor, below which superconductivity vanishes, is $d_{\rm
  S}=5.18\xi_{\rm S}$, which is much larger than in the previous
situations.  We now fix the layer thickness of the superconductor to
$d_{\rm S}=5.52\xi_{\rm S}$ so that $T_c=0.54T_{c,0}$ for $d_{\rm
  F}\to\infty$.  When analysing $T_c$ as a function of $d_{\rm F}$ we
find a completely different behaviour than before: for $d_{\rm F}=0$
the order parameter at the pair-breaking (specular) surfaces,
$x=-d_{\rm S}$ and $x=0$, must be zero.  This suppression of
superconductivity leads to a vanishing critical temperature of the
bi-layer for $d_{\rm F}<0.06\xi_{\rm F}$. The pair-breaking at $x=0$
is weakened when the ferromagnetic layer thickness increases, and for
$d_{\rm F}>0.06\xi_{\rm F}$ the critical temperature becomes finite.
When further increasing $d_{\rm F}$ the critical temperature is
oscillating as in the previous cases, but now starting with a maximum
at $d_{\rm F}=0.15\xi_{\rm F}$; the first minimum can be found at
$d_{\rm F}=0.39\xi_{\rm F}$, and the second maximum at $d_{\rm
  F}=0.62\xi_{\rm F}$.  The difference between the first two maxima is
$0.47\xi_{\rm F}$ as for the unrotated order parameter. It is not
surprising that the periodicity is not affected by the rotation of the
order parameter because the oscillation of the pairing function is an
exclusive result of the ferromagnetic exchange energy $I$, and is
independent of details inside the superconductor. For an infinite
non-magnetic metal layer ($I=0$) the critical temperature of the
rotated $d$-wave superconductor remains unchanged (see \fref{Tc-dS}),
as the suppression of superconductivity is not dominated by the metal
layer but by the surface of the superconductor at $x=-d_{\rm S}$ as
discussed above.  The critical temperature shows no
oscillations with the thickness of the non-magnetic layer (see
\fref{Tc-dF}).

The density of states at $x=d_{\rm F}$ shows a clearly different
behaviour.  In particular for $d_{\rm F}=0$ zero-energy Andreev bound
states exist at the surfaces due to the sign change of the order
parameter for scattered quasi-particles.  For the first maximum of
$T_c$ ($d_{\rm F}=0.15\xi_{\rm F}$) the density of states is
suppressed below the normal state value (see \fref{dosd45}) but a
remainder of the zero-energy bound state is still observable.  For the
first minimum of $T_c$ ($d_{\rm F}=0.39\xi_{\rm F}$) the density of
states for small energies ($E\approx 0$) is enhanced compared to the
normal state value, and for the second maximum of $T_c$ ($d_{\rm
  F}=0.62\xi_{\rm F}$) it is suppressed, which, however, can hardly be
seen in \fref{dosd45}. It is remarkable that the density of states in
the gap region has more structure here than in the previous case. This
is due to the strong angular dependence of the order parameter close
to those directions which are perpendicular to the interface.
Altogether, for an $45^\circ$-rotated $d$-wave order parameter, we
also find that the minima of $T_c$ are related to an enhanced density
of states in the sub-gap region, and vice versa for the maxima of
$T_c$.

\section{Conclusion}

We have studied superconductor-ferromagnet bi-layers for $s$-wave and
$d$-wave superconductors.  In all discussed cases we observed an
oscillating behaviour of the density of states as well as of the
critical temperature when varying the thickness of the ferromagnetic
layer. The origin of these oscillations is the exchange field in the
ferromagnetic metal which leads to a
Larkin-Ovchinnikov-Fulde-Ferrell-like state with a spatially
oscillating pairing function.  In particular we find that the density
of states in the ferromagnetic layer is enhanced in the gap region
when its thickness leads to a minimum of the critical temperature.
When the critical temperature has a maximum the density of states in
the ferromagnetic layer has a gap-like structure.

Comparing the different order parameter symmetries, we observe a
similar behaviour for the $s$-wave and the unrotated $d$-wave cases.
The critical temperature as a function of the ferromagnetic layer
thickness, $d_{\rm F}$, decreases for small values of $d_{\rm F}$
($d_{\rm F}< 0.2\xi_{\rm F}$) and shows oscillations around an
asymptotic value of $T_c$ when further increasing $d_{\rm F}$. In the
$s$-wave case these findings are in agreement with previous
theoretical studies~\cite{Kh97,Ta98}, and are also experimentally
confirmed~\cite{Mu96,Ru02,Si03}.

This behaviour is considerably modified if the $d$-wave superconductor
is rotated by $45^\circ$ with respect to the surface. In this case
superconductivity may even vanish for very thin ferromagnetic layers.
If $d_{\rm F}$ exceeds a critical value, superconductivity can be restored
and oscillations around the asymptotic value of $T_c$ are observed.
This difference in behaviour is due to the sign change of the order
parameter for quasi-particles which are scattered at the surface,
which leads to a suppression of superconductivity. It would be most
interesting to check our results for $d$-wave superconductors also
experimentally.

\section*{Acknowledgements}
We would like to thank P.\ Schwab, L.\ R.\ Tagirov and R.\ Tidecks for
stimulating discussions. This work was supported by the German
Research Foundation (DFG).

\end{document}